\documentclass[pra, twocolumn, showpacs, superscriptaddress,longbibliography]{revtex4-2}

\usepackage{amsmath, hyperref, braket, amssymb, natbib, graphicx,float}

\newcommand{\diff}{\ensuremath{\mathrm{d}}}
\newcommand{\EXP}[1]{\ensuremath{\mathrm{e}^{#1}}}

\newcommand{\dB}{\ensuremath{~\mathrm{dB}}}

\newcommand{\ie}{i.e.~}

\newcommand{\refsec}[1]{Sec.~\ref{#1}}
\newcommand{\refappen}[1]{Appendix.~\ref{#1}}
\newcommand{\refeq}[1]{Eq.~(\ref{#1})}
\newcommand{\Refeq}[1]{(\ref{#1})}
\newcommand{\reffig}[1]{Fig.~\ref{#1}}

\begin{document}


\title{Unconditional measurement-based quantum computation with optomechanical continuous variables}

\author{Oussama Houhou}
\email{houhou.oussama@univ-medea.dz}
\email{o.houhou@hotmail.com}
\affiliation{Laboratory of Physics of Experimental Techniques and Applications, University of M\'ed\'ea, M\'ed\'ea 26000, Algeria}
\affiliation{School of Mathematics and Physics, Queen's University Belfast, BT7 1NN, UK}

\author{Darren W. Moore}
\email{darren.moore@upol.cz}
\affiliation{Department of Optics, Palack\'{y} University, 17. listopadu 1192/12, 771 46 Olomouc, Czech Republic}
\affiliation{School of Mathematics and Physics, Queen's University Belfast, BT7 1NN, UK}

\author{Sougato Bose}
\email{s.bose@ucl.ac.uk}
\affiliation{Department of Physics and Astronomy, University College London, London WC1E 6BT, UK}

\author{Alessandro Ferraro}
\email{a.ferraro@qub.ac.uk}
\affiliation{School of Mathematics and Physics, Queen's University Belfast, BT7 1NN, UK}


\begin{abstract}
	Universal quantum computation encoded over continuous variables can be achieved via Gaussian measurements acting on entangled non-Gaussian states. However, due to the weakness of available nonlinearities, generally these states can only be prepared conditionally, potentially with low probability. Here we show how universal quantum computation could be implemented unconditionally using an integrated platform able to sustain both linear and quadratic optomechanical-like interactions. Specifically, considering cavity opto- and electro-mechanical systems, we propose a realisation of a driven-dissipative dynamics that deterministically prepares the required non-Gaussian cluster states --- entangled squeezed states of multiple mechanical oscillators suitably interspersed with cubic-phase states. We next demonstrate how arbitrary Gaussian measurements on the cluster nodes can be performed by continuously monitoring the output cavity field. Finally, the feasibility requirements of this approach are analysed in detail, suggesting that its building blocks are within reach of current technology. 
\end{abstract}

\maketitle


\section{Introduction}

Measurement-based quantum computation (MBQC) is a powerful approach to process information encoded in quantum systems \cite{briegel2009measurement}, which requires solely local measurements on an entangled state (\textit{cluster state}) \cite{raussendorf2001one, raussendorf2003measurement}. This approach gives significant theoretical insights into fundamental questions about the origin of the power of quantum  computing \cite{van2006universal, gross2009most, bremner2009random, anders2009computational, raussendorf2013contextuality, bermejo2017contextuality}, and it offers promising applicative opportunities provided large enough clusters can be built, including the demonstration of quantum computational supremacy \cite{hangleiter2018anticoncentration,bermejo2018architectures} and the realisation, in condensed matter systems \cite{brennen2008measurement, cai2010universal, li2011thermal, wei2011affleck, aolita2011gapped, else2012symmetry, wei2015universal, miller2015resource}, of fault-tolerant processors with high resilience thresholds \cite{raussendorf2007fault, raussendorf2007topological}.

In view of the relevance of MBQC, major efforts have been devoted to its experimental implementation. In the setting of finite-dimensional (discrete-variable) quantum systems, various experimental demonstrations of small-size MBQC have been reported \cite{walther2005experimental, kiesel2005experimental, prevedel2007high, tame2007experimental, lu2007experimental, vallone2008active, barz2012demonstration, lanyon2013measurement, bell2014experimental,larsen2019deterministic,asavanant2019generation,wu2020quantum,fukui2020temporal,larsen2021deterministic,larsen2021fault,asavanant2021time}. However, the largest clusters to date have been generated in the context of continuous-variable (CV) systems \cite{Braunstein:05, serafini2017quantum}, with photonic clusters composed of up to one million modes \cite{rigas2012generation,yokoyama2013ultra, chen2014experimental, roslund2014wavelength, yoshikawa2016invited, cai2017multimode,larsen2019deterministic,asavanant2019generation}. Such achievements stem from the fact that these clusters belong to the class of Gaussian states \cite{ferraro2005gaussian, Weedbrook:12, adesso2014continuous, genoni2016conditional}, and CV Gaussian entanglement is generally available unconditionally (deterministically), contrary to discrete-variable systems whose entanglement typically relies on post-selection \footnote{See Refs. \cite{lindner2009proposal, schwartz2016deterministic, pichler2017universal} for progresses towards deterministic generation of discrete variable clusters and Refs. \cite{spiller2006quantum, anders2010ancilla, roncaglia2011sequential, proctor2017ancilla, gallagher2018relative} for alternative measurement-based approaches.}. Despite such remarkable progress, in order to realise universal computation \cite{lloyd98computation}, these photonic clusters have to be either equipped with non-Gaussian measurements \cite{menicucci2006universal, gu2009quantum} or interspersed with non-Gaussian states \cite{menicucci2014fault}. Unfortunately, both strategies require high order non-linearities, which are hard to implement deterministically in optics and in fact stand as a major roadblock \footnote{Notice that much effort has been devoted to counteract this issue, leading to proposals in which the necessary non-Gaussian elements can be obtained on-demand \cite{marek2011deterministic, yukawa2013emulating, marshall2015repeat, miyata2016implementation, marek2017general, takeda2017universal}; however these still require the use of quantum memories which are hard to realise \cite{lvovsky2009optical, yoshikawa2013creation} and that we avoid here.}. As a remedy, we propose here to use non-linear optomechanical systems, with the aim of providing a feasible path to unlock the full potential of unconditional MBQC.

Our approach is motivated by recent experimental breakthroughs in cavity optomechanics \cite{milburn2011introduction, aspelmeyer2014cavity}, which lends itself as a disruptive new platform for CVs in which the information carrier is embodied in the centre of mass motion of a mechanical oscillator. Indeed ground state cooling \cite{Connell:2010, teufel:2011, noguchi2016ground,delic2020cooling,magrini2021real,seis2021ground,streltsov2021ground}, squeezing beyond the parametric limit \cite{lecocq2015quantum, wollman2015quantum, pirkkalainen2015squeezing,lei2016quantum,sonar2018strong}, two-oscillator entanglement \cite{ockeloen2018stabilized, riedinger2018remote, barzanjeh2019stationary} and non-locality \cite{marinkovic2018optomechanical} have been achieved experimentally, with further scalability and integrability within reach \cite{massel2012multimode, damskagg2016dynamically, grass2016optical, nielsen2017multimode, noguchi2016strong}. Crucially, optomechanics has a significant advantage to photonics in the unconditional non-linearity embedded in the radiation pressure dynamics \cite{bhattacharya2008optomechanical, thompson2008strong}. For driven systems this manifests primarily as a quadratic coupling in the position of the oscillator \cite{thompson2008strong, woolley2008nanomechanical, hertzberg2010back, rocheleau2010preparation, nunnenkamp2010cooling, purdy2010tunable, sankey2010strong,Hill:11, karuza2012tunable,flowers2012fiber,li2012proposal,hill2013nonlinear,doolin2014nonlinear, kaviani2015nonlinear, paraiso2015position,lee2015multimode,kim2015circuit,abdi2015entangling, kalaee2016design, fonseca2016nonlinear,brawley2016nonlinear,leijssen2017nonlinear,zhang2017photon,dellantonio2018quantum,cattiaux2020beyond,bullier2020quadratic}.

Here we consider a driven-dissipative opto-mechanical system. By taking advantage of the control over the mechanical state granted by externally driving the cavity, and arranging for either dissipative engineering \cite{clerk2013squeezing, Yamamoto:13, houhou2015generation} or continuous monitoring \cite{clerk2008back, genoni2016conditional, darren2017arbitrary}, we are able to provide schemes for the deterministic preparation of non-Gaussian cluster states and local measurements sufficient to achieve computational universality. The integration of these schemes into a single experimental platform constitutes, as far as we know, the first proposal for universal MBQC with CVs that can be implemented unconditionally.

This paper is organised as follows. In \refsec{sec:mbqc} we review measurement-based quantum computation, where we discuss the \emph{standard} method for accomplishing universal quantum computation, and introduce our approach in order to achieve universality. And in \refsec{sec:optomech-implementation} we present the optomechanical system model that will host the resource state suitable for universal quantum computation. This system is driven by a time-dependent multi-tone classical field, and is dissipating to its environment. Then in \refsec{sec:cubic-phase-state} we demonstrate our protocol for preparing a non-Gaussian state, the so-called qubic phase state. This latter is hosted in a mechanical oscillator degrees of freedom. This target state is obtained dissipatively as the steady state of the driven optomechanical system proposed in previous section. Moreover, in \refsec{sec:non-gaussian-cluster} we generalise the proposed protocol of the preceding section in order to generate a non-Gaussian cluster state, composed of nodes of squeezed and cubic phase states, with arbitrary size and geometry. In \refsec{sec:local-gaussian-measurements} we focus on how to perform given measurements on individual mechanical modes, \ie local operations. Specifically, the measurements are Gaussian only, since it is, with the already prepared non-Gaussian cluster state, sufficient to carry out universal computations. Following that, we discuss in \refsec{sec:experimental-feasibility} the feasibility of our proposed scheme, \ie preparing the target state and locally measuring quadratures, in current and near-future experiments. Then a conclusion is given in \refsec{sec:conclusion}. Furthermore, several appendices follow where we study the validity of the approximations used in our derivations (\refsec{sec:validity-rwa}), stability of the system (\refsec{sec:stability}), time scale to prepare cluster state (\refsec{sec:time-scale-cluster}), and an analysis of the effects of the unwanted thermal noise on the quality of the target state (\refsec{sec:noise-cluster}). In addition, we give an example of the preparation of a two-mode cluster (\refsec{sec:two-modes-cluster}) and a demonstration of a cubic phase gate with this latter (\refsec{sec:cubic-phase-gate}).


\section{Measurement-based quantum computation with CVs}\label{sec:mbqc}

\begin{figure}[hbt]
	\includegraphics[scale=1.0]{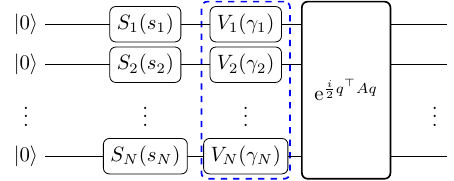}
	\caption{Circuit representing the preparation of a cluster state with squeezing $s$ and adjacency matrix $A$. In the absence (presence) of cubic operations ---given in the dashed box--- the standard (non-Gaussian) cluster is obtained.}
	\label{fig:cluster-state-cubic}
\end{figure}

As said, MBQC is predicated on the existence of a highly entangled multipartite resource state known as \textit{the cluster state}. For our purposes, a cluster state is associated with a mathematical lattice graph $G_{(\mathcal{V,E})}$ of vertices $j\in\mathcal{V}$, and edges $(j,k)\in\mathcal{E}$ that define the adjacency matrix $A$ with entries $A_{j,k}=1$ if $(j,k)\in\mathcal{E}$ and $A_{j,k}=0$ otherwise (with $1\le j,k \le N$). Consider an $N$-oscillator system, with each oscillator $j$ characterised by the canonical position $q_j=\frac{1}{\sqrt{2}}(b_j+b_j^\dagger)$ and momentum $p_j=\frac{1}{i\sqrt{2}}(b_j-b_j^\dagger)$ operators, $b_j$ being their respective annihilation operator. The CV cluster state \cite{zhang2006continuous, menicucci2006universal} is operationally defined by first preparing all vertices (embodied by the oscillators) in a product state of momentum-squeezed vacua $S(s)\ket{0}$, where $\ket{0}=\ket{0}_1\otimes\cdots\otimes\ket{0}_N$, $S(s)=\bigotimes_j S_j(s_j)$, $S_j(s_j)=\exp[\frac{-i}{2}\ (q_jp_j+p_jq_j) \ln s_j]$, and $s\equiv (s_1,\ldots,s_N)$ is a shorthand for the degree of squeezing. Then, controlled-phase operations $\mathrm{CZ}_{jk}=\EXP{iq_jq_k}$ are applied for any edge $(j,k)\in\mathcal{E}$. These can be compactly written defining the multi-oscillator operator $E(A)=\EXP{\frac{i}{2}q^\top Aq}$, with $q=(q_1,\ldots,q_N)^\top$. Consequently, the resulting \textit{standard} cluster state is given by $\ket{s,A}=E(A)S(s)\ket{0}$ (see \reffig{fig:cluster-state-cubic}); this is a Gaussian state and the degree of squeezing $s$ (with $s\ge 1$ for momentum-squeezing) determines its quality for computational purposes \cite{gu2009quantum}.

The computation proceeds via a series of local projective measurements on the cluster nodes. These measurements implement the gates of the program to be computed, whose output is embodied in the state of the non-measured nodes. The Lloyd-Braunstein criterion \cite{lloyd98computation}, first developed for circuit-based computation, allows a distinction to be drawn between Gaussian and non-Gaussian gates. A finite set of Gaussian gates is sufficient to perform any multimode Gaussian operation. However, it is only when an additional non-Gaussian gate is at disposal that universality is unlocked, in the sense that any Hamiltonian can be simulated to arbitrary precision. In MBQC, Gaussian measurements on the cluster $\ket{s,A}$ are sufficient to implement arbitrary Gaussian gates \cite{gu2009quantum}, including in extremely compact ways \cite{ferrini2013compact}. On the other hand, as mentioned, several proposals for implementing non-Gaussian gates are extant in the literature \cite{ghose2007non, gu2009quantum, marek2011deterministic,yukawa2013emulating, marshall2015repeat, miyata2016implementation, takeda2017universal, marek2017general}. Here we focus on a method in which the standard cluster is modified using non-Gaussian resources --- called \textit{cubic-phase states} \cite{gottesman2001encoding}. This modified non-Gaussian cluster is particularly advantageous for scaling to large numbers of operations since it allows for the measurement strategy to remain Gaussian \cite{gu2009quantum, gottesman2001encoding}.

We will first present a general exposition of the optomechanics model we wish to base our proposal on, and then introduce two complementary schemes allowing us to prepare the \textit{modified} non-Gaussian cluster and perform on it arbitrary Gaussian measurements. 


\section{Optomechanics implementation}\label{sec:optomech-implementation}

Consider an array of $N$ mechanical resonators, each with distinct frequency $\Omega_j$, immersed in a cavity field with annihilation operator $a$ and frequency $\omega$ and driven by a time-dependent external field $\epsilon(t)$. The Hamiltonian for such a system is
\begin{equation}
	H=\omega a^\dagger a+\sum_{j=1}^N\Omega_j b^\dagger_jb_j+\epsilon(t)^*a+\epsilon(t)a^\dagger\,.
\end{equation}
Due to radiation pressure the cavity frequency becomes dependent on the mechanical positions $q_j$ \cite{aspelmeyer2014cavity}. We expand $\omega$ in powers of $q_j$ up to the second order:
\begin{equation}
	\omega=\omega_c+\sum_{j=1}^N\left(g_L^{(j)}q_j+g_Q^{(j)}q_j^2+\cdots\right)\,,
\end{equation}
with $g_{L}^{(j)}=\frac{\partial\omega}{\partial q_j}$ and $g_Q^{(j)}=\frac12\frac{\partial^2\omega}{\partial q_j^2}$ the position and position-squared couplings of the $j^\text{th}$ mechanical oscillator with the cavity field. In addition, we consider the case of a multi-tone drive,
\begin{equation}
	\epsilon(t)=\sum_k\epsilon_k e^{-i\omega_kt}\,,
\end{equation}
with $\epsilon_k$ the complex driving amplitudes and $\omega_k$ the driving frequencies. The standard \textit{linearisation} procedure for an externally driven cavity \cite{aspelmeyer2014cavity} may be expanded to include the multiple mechanical modes, the multi-tone drive and the position-squared coupling as follows. Allowing our system to be in contact with a vacuum reservoir for the cavity and a thermal bath for the mechanical oscillators leads to the following Heisenberg~-~Langevin equations \cite{Gardiner:00} for the system operators:
\begin{align}
	\dot{q_j}(t)	&=	\Omega_j p_j\,,\label{eqn:eqs-motion-q-j}\\
	\dot{p_j}(t)	&=	-\Omega_j q_j-a^\dagger a\left(g_L^{(j)}+2g_Q^{(j)} q_j\right)\nonumber\\
					&\quad	-\Gamma_j p_j+\xi_j(t)\,,\label{eqn:eqs-motion-p-j}\\
	\dot{a}(t)		&=	\left(-\frac{\kappa}{2}-i\omega_c\right)a-ia\sum\limits_{j=1}^N\left(g_L^{(j)} q_j+g_Q^{(j)} q_j^2\right)\nonumber\\
					&\quad	-i\epsilon(t)+\sqrt{\kappa}\ a_{\text{in}}\,,\label{eqn:eqs-motion-a}
\end{align}
where $\kappa$ and $\Gamma_j$ are the damping rates for the cavity mode and the $j^{\text{th}}$ mechanical oscillator, and $a_{\text{in}}$ and $\xi_j$ are the input noise operators for the cavity and mechanical oscillator respectively, satisfying the correlation relations:
\begin{align}
	\braket{a_{\text{in}}^\dagger (t) a_{\text{in}}(t')}	&=	0\,,\\
	\braket{a_{\text{in}}(t) a_{\text{in}}^\dagger (t')}	&=	\delta(t-t')\,,\\
	\braket{\xi_j^\dagger (t)\xi_j(t')}	&=	\bar{n}_j \delta(t-t')\,,\\
	\braket{\xi_j(t)\xi_j^\dagger (t')}	&=	(\bar{n}_j +1)\ \delta(t-t')\,,
\end{align}
with $\bar{n}_j$ denoting the mean phonon number.

We aim to derive an effective Hamiltonian for the system involving quantum fluctuations around the (classical) fields steady states. Replacing the system operators, in equations \Refeq{eqn:eqs-motion-q-j}--\Refeq{eqn:eqs-motion-a}, by their mean-fields: $\braket{a}\equiv\alpha$, $\braket{q_j}\equiv Q_j$ and $\braket{p_j}\equiv P_j$, the classical equations of motion become:
\begin{align}
	\dot{Q}_j(t)		&=	\Omega_j P_j\,,\\
	\dot{P}_j(t)		&=	-\Omega_j Q_j-|\alpha|^2\left(g_L^{(j)}+2g_Q^{(j)} Q_j\right)-\Gamma_j P_j\,,\label{eqn:P-j-dot}\\
	\dot{\alpha}(t)		&=	\left(-\frac{\kappa}{2}-i\left[\omega_c+g_L^{(j)} Q_j+g_Q^{(j)} Q_j^2\right]\right)\alpha\nonumber\\
						&\quad	-i\epsilon(t)\,.
\end{align}
We consider the following ansatz for the intra-cavity field at the steady state \cite{milburn2011introduction}:
\begin{equation}\label{eqn:ansatz-alpha}
	\alpha=\sum\limits_k \alpha_k\EXP{-i\omega_k t}\,,
\end{equation}
where the constants $\alpha_k$ are the complex amplitudes of the cavity at the steady state. By substituting expression~\Refeq{eqn:ansatz-alpha} in \refeq{eqn:P-j-dot} we find:
\begin{align}
	\dot{P}_j(t)=&-\Omega_j Q_j-\Gamma_j P_j\nonumber\\
	&-\left(g_L^{(j)}+2g_Q^{(j)} Q_j\right)\sum\limits_{k,\ell}\alpha_k^*\alpha_\ell\EXP{i(\omega_k-\omega_\ell)t}\,.\label{eqn:P-j-dot-2}
\end{align}
If we assume weak coupling such that for $k\ne \ell$ we have
\begin{equation}
	\left|g_{L,Q}^{(j)}\ \alpha_k\alpha_\ell\right|\ll\Omega_j\,,
\end{equation}
the time dependent terms in \refeq{eqn:P-j-dot-2} can be neglected. And if we denote by $Q_j^{(0)}$ and $P_j^{(0)}$ the values of position and momentum at the steady state, it is easy to find the following:
\begin{align}
	P_j^{(0)}	&=	0\,,\\
	Q_j^{(0)}	&=	\frac{-g_L^{(j)}\sum\limits_k|\alpha_k|^2}{\Omega_j+2g_Q^{(j)} \sum\limits_k|\alpha_k|^2}\,,\\
	\alpha_k	&=	\frac{-i\epsilon_k}{\frac{\kappa}{2}+i\left(-\Delta_k+g_L Q_0+g_Q Q_0^2\right)}\,,
\end{align}
where $\Delta_k\equiv\omega_k-\omega_c$ is the detuning of the $k^{\text{th}}$ drive with respect to the cavity.
	
Having obtained the steady state for all fields, we can derive a Hamiltonian of the system in terms of the quantum fluctuations around the classical steady state values. First, we split the system operators into a classical part (denoted as $\alpha_k$, $Q^{(0)}_j$ and $P^{(0)}_j$) and quantum fluctuations (denoted with a slight abuse of notation as $a$, $q_j$ and $p_j$),
\begin{equation}\label{eqn:split-classical-quantum}
	\left.
	\begin{aligned}
		a&\rightarrow a+\sum\limits_k\alpha_k\EXP{-i\omega_k t}\,,\\
		q_j&\rightarrow q_j+Q^{(0)}_j\,,\\
		p_j&\rightarrow p_j+P^{(0)}_j\,,
	\end{aligned}
	\right\}
\end{equation}
then we substitute \Refeq{eqn:split-classical-quantum} in equations~\Refeq{eqn:eqs-motion-q-j}--\Refeq{eqn:eqs-motion-a}. Assuming a strong drive, $\alpha_k\gg 1$, we find:
\begin{widetext}
	\begin{align}
		\dot{q}_j	&=	\Omega_j p_j\,,\label{eqn:q-j-dot-lin-a}\\
		\dot{p}_j	&\approx	-\Omega_j q_j-\left(a^\dagger\sum\limits_k\alpha_k\EXP{-i\omega_k t}\ +a\sum\limits_k\alpha_k^*\EXP{i\omega_k t}\right)\left(g_L^{(j)}+2g_Q^{(j)}\ Q_j^{(0)}+2g_Q^{(j)}\ q_j\right)-\Gamma_j p_j+\xi_j(t)\,,\label{eqn:p-dot-lin-a}\\
		\dot{a}		&\approx	\left(\frac{-\kappa}{2}-i \omega_c\right)a-\sum\limits_k i\alpha_k\EXP{-i\omega_k t}\left(\left[g_L^{(j)}+2g_Q^{(j)}\ {Q_j^{(0)}}^2\right]q_j+g_Q^{(j)}\ q_j^2\right)+\sqrt\kappa\ a_{\text{in}}(t)\,.\label{eqn:a-dot-lin-a}
	\end{align}
\end{widetext}
Equations \Refeq{eqn:q-j-dot-lin-a}--\Refeq{eqn:a-dot-lin-a} correspond to the following effective Hamiltonian:
\begin{widetext}
	\begin{equation}\label{eqn:optomech-hamiltonian-lin}
		H=\omega_c a^\dagger a+\sum\limits_{j=1}^N\Big[\Omega_jb^\dagger_jb_j+\sum\limits_k\left(\alpha_k\EXP{-i\omega_k t} a^\dagger+\alpha_k^*\EXP{i\omega_k t} a\right)\left(\sqrt2\ G_L^{(j)} q_j+2G_Q^{(j)}\ q_j^2\right)\Big]\,,
	\end{equation}
\end{widetext}
where we defined $\sqrt2\ G_L^{(j)}\equiv g_L^{(j)}+2g_Q^{(j)} {Q_j^{(0)}}^2$ and $2G_Q^{(j)}\equiv g_Q^{(j)}$.
	
The explicit time-dependence of Hamiltonian~\Refeq{eqn:optomech-hamiltonian-lin} can be removed by, first, going to a frame rotating with the free terms of the system where the Hamiltonian transforms to
\begin{widetext}
\begin{equation}\label{eqn:optomech-hamiltonian-linear1}
	\mathcal{H}=\sum_{j=1}^N\sum_k\left(\alpha_k\EXP{-i\Delta_k t}a^\dagger+\alpha_k^*\EXP{i\Delta_k t} a\right)\left[G_L^{(j)}\left(b_j\EXP{-i\Omega_jt}+b_j^\dagger\EXP{i\Omega_jt}\right)+G_Q^{(j)}\left(b_j\EXP{-i\Omega_jt}+b_j^\dagger\EXP{i\Omega_jt}\right)^2\right]\,.
\end{equation}
\end{widetext}
Then, we consider four driving fields per each mechanical resonator $j$ with detunings $\Delta^{(j)}_1=-\Omega_j,\ \Delta^{(j)}_2=\Omega_j,\ \Delta^{(j)}_3=-2\Omega_j,\ \Delta^{(j)}_4=2\Omega_j$ and amplitudes $\alpha^{(j)}_\ell$ ($\ell=1,\ldots,4$). Moreover, we consider an additional drive that is resonant with the cavity ($\Delta_5=0$), with amplitude $\alpha_5$. Hamiltonian~\Refeq{eqn:optomech-hamiltonian-linear1}, in the rotating wave approximation (RWA), becomes (see \refappen{sec:validity-rwa})
\begin{multline}\label{eqn:optomech-hamiltonian-N-0}
	\mathcal{H}=a^\dagger\sum_{j=1}^N\left(g^{(j)}_1 b_j+g^{(j)}_2 b_j^\dagger\,+\right.\\\left.+\,g^{(j)}_3 b_j^2+g^{(j)}_4 {b_j^\dagger}^2+g^{(j)}_5\{b_j,b_j^\dagger\}\right)+\text{H.c.}\,,
\end{multline}
with $g^{(j)}_\mu\equiv\alpha^{(j)}_\mu G^{(j)}_L$, $g^{(j)}_\nu\equiv\alpha^{(j)}_\nu G^{(j)}_Q$ ($\mu=1,2$; $\nu=3,4$) and $g_5^{(j)}\equiv\alpha_5 G_Q^{(j)}$ the amplifications of the single phonon-photon couplings due to the external driving. Notice that independent control over each term in the Hamiltonian~\Refeq{eqn:optomech-hamiltonian-N-0} is possible \footnote{Note that the parameters $g_\ell^{(j)}$, $\ell=1,\ldots,4$, can be tuned by varying both the bare optomechanical couplings and the driving amplitudes, while $g_5^{(j)}$ is set by changing $G_Q^{(j)}$ only.}, which is in turn crucial for our purposes. The aforementioned RWA holds in a regime satisfying $|\alpha_\ell^{(j)} G_\sigma^{(k)}|\ll\Omega_j$ and $|\alpha_5G_\sigma^{(j)}|\ll\Omega_j$ ($j,k=1,\ldots,N$, $\ell=1,\ldots,4$, $\sigma=\mathrm{L,Q}$), given that the frequencies $\Omega_j$ do not overlap, see \refappen{sec:validity-rwa}.

As said, dissipation is central for our aims. We model the evolution of the system by a master equation in which the cavity mode dissipates at a rate $\kappa$ and the mechanical oscillators are in contact with a thermal bath \cite{Carmichael:93,Gardiner:00}:
\begin{multline}\label{eqn:master-equation}
	\dot{\rho}(t)=-i[\mathcal{H}, \rho(t)]+\kappa D[a]\rho(t)\\ +\sum_{j=1}^N\Gamma_j(\bar{n}_j+1)D[b_j]\rho(t)+\Gamma_j\bar{n}_jD[b_j^\dagger]\rho(t)\,,
\end{multline}
where the standard super-operator for Markovian dissipation is denoted as $D[f]\rho=f\rho f^\dagger-\frac12\{f^\dagger f,\rho\}$ ($f=a,b_j$).


\section{The cubic phase state}\label{sec:cubic-phase-state}

In this section, we set $N=1$ and omit all subscripts/superscripts related to the oscillator.

The finitely-squeezed cubic phase state of a single oscillator is defined as \cite{gottesman2001encoding}
\begin{equation}\label{eqn:cubic-phase-state-2}
	\ket{\gamma, s}=\EXP{i\gamma q^3}S(s)\ket{0}\,.
\end{equation}

A core result of our proposal is that the cubic phase state of a single mechanical oscillator can be unconditionally generated as the steady state of the dynamics given in \refeq{eqn:master-equation} (with $N=1$), applying suitable drive amplitudes and phases. Setting $N=1$ the Hamiltonian~\Refeq{eqn:optomech-hamiltonian-N-0} simplifies to
\begin{equation}\label{eqn:hamilt-1}
	H_\text{cub}=a^\dagger\left(g_1 b+g_2 b^\dagger+g_3 b^2+g_4 {b^\dagger}^2+g_5\{b,b^\dagger\}\right)+\text{H.c.}\,.
\end{equation}
The coefficients of the linear terms, $g_1$ and $g_2$, are associated only with Gaussian steady states \cite{houhou2015generation}. Indeed, the ratio of the amplitudes of these determines the degree of squeezing \footnote{We describe the level of squeezing of the state given in \refeq{eqn:cubic-phase-state-2} as $10\log_{10}s^2$~dB \cite{gu2009quantum}.} of the steady state \cite{houhou2015generation, clerk2013squeezing}. Non-Gaussianity at the steady state derives instead from the remaining coefficients as follows. By choosing the driving strengths as $g_2=-r g_1$, $g_3=g_4=g_5=\frac{-3i}{2\sqrt2}\ \gamma(1+r)g_1$, with $r=\frac{s^2-1}{s^2+1}$, we obtain the Hamiltonian 
\begin{multline}
	H_\text{cub}=g_1 a^\dagger\Big(b-r b^\dagger\\
	-\frac{3i \gamma}{2\sqrt2}(1+r)(b+b^\dagger)^2\Big)+\text{H.c.}\,,
\end{multline}
which also can be put in the form
\begin{equation}\label{eqn:H_cubic}
	H_\text{cub}=g_1\sqrt{1-r^2}\ a^\dagger UbU^\dagger + \text{H.c.}\,,
\end{equation}
where $U=\EXP{i\gamma q^3}\EXP{-\frac{i}{2}\ln s\ (qp+pq)}$. When neglecting the mechanical thermal noise, \ie $\Gamma=0$, the master equation~\Refeq{eqn:master-equation} can be rewritten as
\begin{equation}\label{eqn:master-equation-transformed}
	\dot{\tilde{\rho}}(t)=-i[\tilde{H}, \tilde{\rho}(t)]+\kappa D[a]\tilde{\rho}(t)\,,
\end{equation}
where we defined $\tilde{\rho}\equiv U^\dagger\rho U$ and
\begin{equation}
	\tilde{H}\equiv U^\dagger H_\text{cub}U=g_1\sqrt{1-r^2}\ a^\dagger b + \text{H.c.}\,.
\end{equation}
Notice that the new transformed Hamiltonian is a beam-splitter-like interaction. Therefore, the steady state of the dynamics described by the new master equation~\Refeq{eqn:master-equation-transformed} is the vacuum for both the cavity and the \textit{new} $U$-transformed mechanical mode. Consequently, the steady state of the system's dynamics governed by the original master equation is the state $\ket{0}_c\otimes\ket{\gamma,s}$ where $\ket{0}_c$ is the vacuum state of the cavity and $\ket{\gamma,s}$ is the mechanical finitely squeezed cubic phase state defined in \refeq{eqn:cubic-phase-state-2}. We stress the fact that the obtained cubic phase state is prepared \textit{deterministically} and this preparation protocol is \textit{independent} of the system's initial conditions. Moreover, the stability condition of the system's dynamics is inherited from the linear system: $0\le r<1$ \footnote{Notice that we always have $s\ge 1$, and since $r=\frac{s^2-1}{s^2+1}$ then we must have $0\le r<1$.}, see \refappen{sec:stability}.
		
\begin{figure}[hbt]
	\begin{center}
		\includegraphics[scale=0.6]{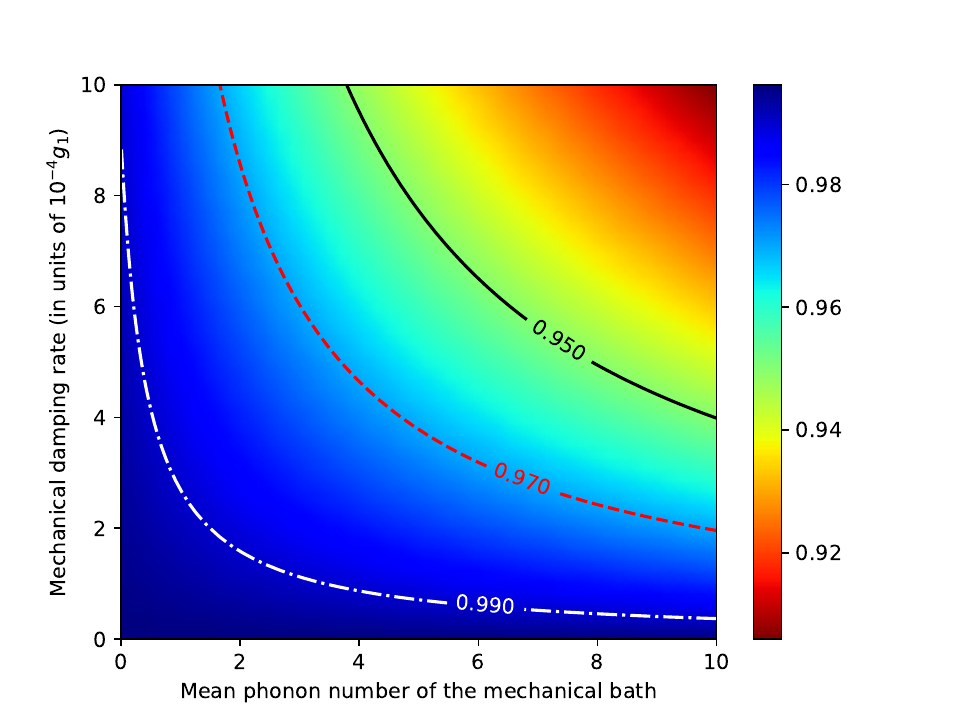}
	\end{center}
	\caption{\label{fig:fidelity-temp-damp}Fidelity of the noisy cubic phase state with the noiseless one as a function of the mean phonon number ($\bar{n}$) and mechanical damping rate ($\Gamma$). Each point of the plot was obtained with the following parameters: $r=0.52$ ($5\dB$), $\gamma=0.2$, and $\kappa=10~g_1$.}
\end{figure}

In order to consider the effect of non-zero mechanical noise, we numerically find the steady state of \refeq{eqn:master-equation} and then we calculate the fidelity between the latter and the state in \refeq{eqn:cubic-phase-state-2}. This is shown in \reffig{fig:fidelity-temp-damp} where we plot the fidelity as a function of the mean phonon number of the bath and the mechanical damping rate. As expected, the mechanical noise has a noxious effect on the target cubic phase state; the higher the temperature (quantified by $\bar{n}$) or mechanical damping rate ($\Gamma$) the lower the fidelity.

This analysis shows that a cubic phase state can be generated in the massive mechanical oscillator of an optomechanics experiment. This is a result of interest in its own, given the highly non classical character of such a state --- which displays a non-positive Wigner function and a high degree of quantum non-Gaussianity \cite{takagi2018convex, albarelli2018resource} --- and its deterministic attainability. As mentioned, for this state to be considered as a resource for computing, we must also show that it can be embedded in a standard Gaussian cluster state \footnote{A seminal proposal to realise, via optomechanical-like interactions, states potentially useful for MBQC is given in Ref.~\cite{pirandola2006continuous}; however, there a non-linearised and probabilistic approach is considered. We also notice that the recent deterministic proposal in Ref.~\cite{brunelli2018unconditional} use similar techniques to the ones presented here but the states obtained there are not useful for MBQC (see also \cite{brunelli2018linear} for further analysis).}.


\section{Non-Gaussian cluster states}\label{sec:non-gaussian-cluster}

We aim to generate a modified non-Gaussian cluster state sufficient to perform universal computation by interspersing the standard state $\ket{s,A}$  with cubic-phase states. In particular we will now show how the dissipative dynamics described by \refeq{eqn:master-equation} can be adapted to generate the state
\begin{equation}
	\ket{\gamma,s,A}=E(A)V(\gamma)S(s)\ket{0}\,,
\end{equation}
where $\gamma\equiv(\gamma_1,\ldots,\gamma_N)$ denotes the cubic non-linearities, and we have defined $V(\gamma)=\bigotimes_{j=1}^N V_j(\gamma_j)$ and $V_j(\gamma_j)=\EXP{i\gamma_jq_j^3}$ (see \reffig{fig:cluster-state-cubic}). The state $\ket{\gamma,s,A}$ allows the implementation of universal computation since it can be composed of nodes with zero non-linearity, as the standard Gaussian one, and nodes with $\gamma_j\neq0$. For any given computation, Gaussian measurements will then ``tailor'' this non-Gaussian cluster accordingly to the program to be implemented. In this way, cubic gates $V(\gamma)$ can be implemented only when needed \footnote{Similar tailoring techniques have been considered in the context of both continuous \cite{menicucci2014fault} and discrete variables \cite{bermejo2017contextuality}. In the latter, Gaussian and non-Gaussian gates are substituted with Clifford and non-Clifford ones.}.

Adapting the Hamiltonian switching scheme considered in Refs.~\cite{Yamamoto:13,houhou2015generation}, one can generate the state $\ket{\gamma,s,A}$ via dissipation engineering. The switching scheme involves $N$ steps such that at each one the driving fields are tuned to implement the transformation
\begin{equation}
	d_k=E(A)V(\gamma)S(s)\ b_k\ \Big(E(A)V(\gamma)S(s)\Big)^\dagger\,.
\end{equation}
This implies that, at the $k^{\text{th}}$ step, the Hamiltonian is
\begin{equation}\label{eqn:H-cluster}
	H_\text{clust}^{(k)}=\beta(a^\dagger d_k+a d_k^\dagger)\,,
\end{equation}
where $\beta>0$ is a parameter proportional to the driving power \footnote{The protocols discussed in this work give a recipe on how to choose the drivings. In particular, the ratios between the drivings are what dictate which target state is to be prepared, whereas the strengths of the drives (described by $\beta$) control the speed of reaching the steady state.}, and the driving strengths $g_i^{(j)}$ appearing in \refeq{eqn:optomech-hamiltonian-N-0} are chosen according to the transformation $b_k\rightarrow d_k$. At each step, the system is allowed to reach its steady state (\ie the vacuum of the collective mode $d_k$) and then the Hamiltonian is switched, by modifying the driving fields, for the next step to begin. Therefore, if the system is initially in vacuum (and neglecting the mechanical damping), after the $N$ steps the mechanical state is given by the target cluster state, in the basis of the local modes $\{b_1,\dots,b_N\}$ \footnote{We should mention that our switching scheme introduced here is not only a generalisation of a previous protocol \cite{houhou2015generation} for the generation of Gaussian cluster states, but also conforms with the canonical preparation of Gaussian cluster states if we restrict our selves to first sideband drivings only.}.

Now we show explicitly how the optomechanical couplings should be chosen in order to obtain the target (non-Gaussian) cluster state. The unitary transformation for the collective modes can be written as
\begin{equation}\label{eqn:d-ell}
	d_\ell=\sum\limits_{j=1}^N\left[\mathcal{R}_{\ell j}\ b_j+\mathcal{S}_{\ell j}\ b_j^\dagger+\mathcal{T}_{\ell j}\left(b_j+b_j^\dagger\right)^2\right]\,,
\end{equation}
where the matrices $\mathcal{R}$, $\mathcal{S}$ and $\mathcal{T}$ are given in terms of the squeezings $s$, cubic nonlinearity parameters $\gamma$ and the target adjacency matrix $A$ as follows:
\begin{align}
	\mathcal{R}	&=	D_+-\frac{i}{2}(D_++D_-)A\,,\\
	\mathcal{S}	&=	-D_--\frac{i}{2}(D_++D_-)A\,,\\
	\mathcal{T}	&=	\frac{-3i}{2\sqrt2}\ D_\gamma(D_++D_-)\,,
\end{align}
where the matrices $D_\pm$ and $D_\gamma$ are given by
\begin{align}
	D_\pm		&=	\frac12\ \text{diag}\left(s_1\pm\frac{1}{s_1},\ldots,s_N\pm\frac{1}{s_N}\right)\,,\\
	D_\gamma	&=	\text{diag}\left(\gamma_1,\ldots,\gamma_N\right)\,.
\end{align}
	
The switching protocol is performed by choosing the optomechanical couplings such that at the $\ell^{\text{th}}$ step we set:
\begin{align}
	g^{(j)}_1	&=	\beta \mathcal{R}_{\ell j}\,,\\
	g^{(j)}_2	&=	\beta \mathcal{S}_{\ell j}\,,\\
	g^{(j)}_3	&=	g^{(j)}_4=g^{(j)}_5=\beta \mathcal{T}_{\ell j}\,,\label{eqn:switch-program-quadratic}
\end{align}
for some constant parameter $\beta$. We should mention here that the switching program requires, at the step $\ell$, to set all the quadratic couplings, except $g_5^{(\ell)}$, to zero (this is clear from \refeq{eqn:switch-program-quadratic} where the matrix $\mathcal{T}$ is diagonal). On the other hand, the parameters $g_5^{(j)}$ are tunable only through the control of the quadratic couplings; $g_5^{(j)}=\alpha_0 G_Q^{(j)}$, and the resonant drive $\alpha_0$ will lead to all the terms $\{b_j,b_j^\dagger\}$ to be resonant in the Hamiltonian. Therefore, for our protocol to work one needs to be able to switch on and off the quadratic couplings $G_Q^{(j)}$ at will in every step of the switching scheme in order to kill all terms $\{b_j,b_j^\dagger\}$ for $j\ne \ell$ at step $\ell$.

The system considered here will always relax to {\em one and only one} steady state, ie. the dynamical system has always one attractor. We should stress the fact that this uniqueness of the steady state is per initial state of the system, ie. for every initial state there corresponds exactly one steady state. One special case is that of a system with one mechanical mode ($N=1$), where all initial states lead to the same steady state, the cubic phase state, as shown in \refsec{sec:cubic-phase-state}. That being said, for a system with more than one mechanical mode ($N\ge 2$), starting from two different initial states, the system will reach two different steady states. In particular, if the dynamics starts from vacuum and we perform Hamiltonian switching, then the system reaches a unique steady state at every step, and the steady state of the last step will be our target cluster state.

\begin{figure}[hbt]
	\begin{center}
		\includegraphics[scale=0.57]{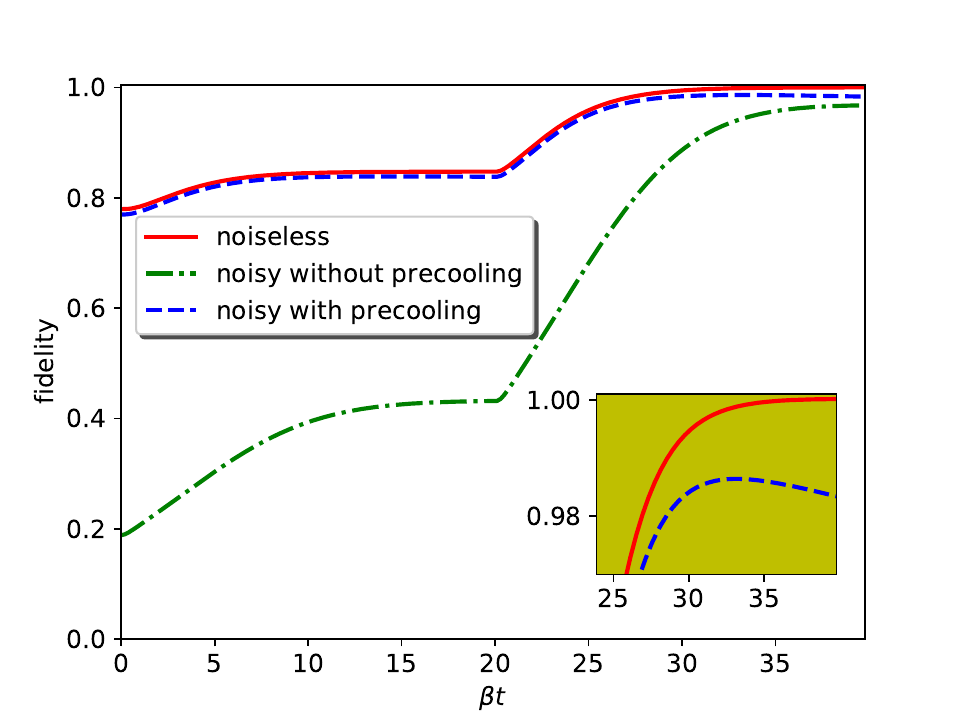}
	\end{center}
	\caption{The fidelity of the preparation of a two-node non-Gaussian cluster state. The nodes of the cluster consist of a squeezed state and a cubic phase state with same amount of squeezing. We used the following parameters: $s_1=s_2=1.78$ ($5\dB$), $\gamma_1=0$, $\gamma_2=0.1$, $\kappa=10~\beta$ and evolution duration $\tau=20~\beta^{-1}$. Pre-cooling (dashed line) the oscillators close to the ground state greatly increases the (maximum) achievable fidelity of the scheme.}
	\label{fig:TwoModeFidelity}
\end{figure}

\reffig{fig:TwoModeFidelity} demonstrates the effectiveness of the switching scheme for generating a two-node non-Gaussian cluster. In the absence of mechanical noise (solid red line), the fidelity with the target state increases monotonically in each step and it reaches unit fidelity at the steady state (at the end of the second step, provided longer evolution time is allowed). When the mechanical environment is considered (dot dashed line), the fidelity reaches a maximum (during the second step) before the noise starts to negatively affect the quality of the target cluster state. As already seen in Fig.~\ref{fig:fidelity-temp-damp}, the thermal noise has a detrimental effect on the performance of the switching scheme, however high fidelities can still be achieved. Part of this negative effect is due to the fact that the oscillators are assumed to be initialised in thermal equilibrium with their environment  (with mean phonon numbers $\bar{n}_1=10$ and $\bar{n}_2=1$ and mechanical damping $\gamma_m=10^{-4}\beta$), rather than in the ground state. This effect can then be circumvented to a large degree by first independently cooling the oscillators (red detuned sideband cooling) \cite{marquardt2007quantum, wilson2007theory}. This can be seen in the dashed blue curve of \reffig{fig:TwoModeFidelity}, which in fact closely approximates the noiseless scenario.

\section{Local Gaussian measurements}\label{sec:local-gaussian-measurements}

For the non-Gaussian cluster state above to be useful in quantum computation, one finally requires the capacity to perform Gaussian measurements on individual nodes. Unfortunately, the mechanical modes embodying our cluster state are not directly accessible to measurement and must be probed instead using the cavity field as a detector. Conveniently, since the resonators are assumed to have distinct and well-spaced frequencies (see \refappen{sec:validity-rwa} for details), we may address each oscillator individually. In particular, by properly driving the system, it is possible to engineer a quantum non-demolition (QND) interaction between the cavity position quadrature and an arbitrary quadrature of any given oscillator \cite{clerk2008back, darren2017arbitrary}.

Consider again Hamiltonian \Refeq{eqn:optomech-hamiltonian-N-0} with $g_3^{(j)}=g_4^{(j)}=g_5=0$ for $j=1,\ldots,N$, addressing only the first sidebands. Let $\left|g_1^{(j)}\right|=\left|g_2^{(j)}\right|=\beta_j$ and $\text{arg}\left(g_2^{(j)}\right)=-\text{arg}\left(g_1^{(j)}\right)\equiv\phi_j$. In this case one has a sum of QND interactions,
\begin{equation}
	H_\text{meas}=2X\sum_{j=1}^N\beta_jQ_{\phi_j}^{(j)}\,,
\end{equation}
where $X=\frac{a+a^\dagger}{\sqrt2}$ is the cavity position quadrature and $Q_{\phi_j}^{(j)}=q_j\cos\phi_j+p_j\sin\phi_j$ is an arbitrary quadrature of the mechanical mode $j$. Each oscillator can be addressed in turn by setting all but the amplitude of interest to zero. In this case we have an $N$-step process with each step described by
\begin{equation}
	H_\text{meas}^{(k)}=2\beta_k X Q_{\phi_k}^{(k)}\,.
\end{equation}
Let us stress that the mechanical quadrature $Q_{\phi_k}^{(k)}$ to be measured, which in turn depends on the program to be implemented, is simply selected by the phase of the external driving. Continuously monitoring, via homodyne detection, the output cavity field's position quadrature drives the mechanical system towards an eigenstate of the chosen quadrature (represented by a vacuum state squeezed along an appropriate axis defined by $\phi_k$). For the purposes of computation, this is equivalent to performing a projective quadrature measurement directly onto the cluster state \cite{darren2017arbitrary}. As said, the latter are in turn sufficient to perform any multimode operation, when operating on the non-Gaussian cluster  $\ket{\gamma,s,A}$.

In \refappen{sec:cubic-phase-gate} we provide an example of how to implement the minimal building block of universal MBQC by using the tools introduced so far. In particular, we consider the universal non-Gaussian gate defined as the operator $V=\EXP{i\gamma q^3}$ \cite{Weedbrook:12} --- called \textit{cubic phase gate} --- and show that it can be reliably implemented on a squeezed state via local Gaussian measurements on the two-node non-Gaussian cluster of Fig.~\ref{fig:TwoModeFidelity}.


\section{Experimental feasibility}\label{sec:experimental-feasibility}

The protocol proposed above to prepare non-Gaussian cluster states requires physical platforms exhibiting linear and quadratic position coupling with the cavity field. Moreover, the system needs to operate in the resolved sideband regime and the conditions $|\alpha_\ell^{(j)} G_\sigma^{(k)}|\ll\Omega_j$ and $|\alpha_5G_\sigma^{(j)}|\ll\Omega_j$ ($j,k=1,\ldots,N$, $\ell=1,\ldots,4$, $\sigma=\text{L,Q}$) must be met to ensure the validity of the RWA used in our derivation of the dynamics. These requirements may be realised in current and near future experiments. In fact, there are many platforms that can be used to implement our scheme, including membrane-in-the-middle configurations \cite{thompson2008strong, sankey2010strong, flowers2012fiber, karuza2012tunable, lee2015multimode,hauer2018phonon,dalafi2018effects,zhu2018controllable}, ultracold atoms inside a cavity \cite{purdy2010tunable}, photonic crystals \cite{paraiso2015position, kalaee2016design, leijssen2017nonlinear, kaviani2015nonlinear}, circuit-QED \cite{kim2015circuit}, electro-mechanical systems \cite{woolley2008nanomechanical, hertzberg2010back, rocheleau2010preparation, massel2012multimode, dellantonio2018quantum, cattiaux2020beyond}, micro-disks \cite{li2012proposal,Hill:11, hill2013nonlinear, doolin2014nonlinear}, and optically levitated particles \cite{chang2010cavity, abdi2015entangling, fonseca2016nonlinear,bullier2020quadratic,delic2020levitated,he2020normal}. In particular, very large quadratic couplings are within reach of current experiments \cite{Hill:11, paraiso2015position, kalaee2016design, gu2019generation, sainadh2020displacement}. Also we mention that linear to quadratic ratios of up to $10^2$ may be obtained \cite{zhang2017photon, brawley2016nonlinear}.

Furthermore, the linear to quadratic couplings ratio can be improved by optimising the experimental design. For instance, one may exploit the membrane tilting in membrane-in-the-middle setups \cite{thompson2008strong,sankey2010strong} or fine positioning the microdisc in microtorid optomechanical systems \cite{li2012proposal}. Also, our protocols can be implemented in electrical circuits by controlling the bias flux and coupling capacitance as proposed in \cite{kim2015circuit}, or considering magnetically or optically levitated particles as suggested in \cite{pino2018chip, chang2010cavity}.

Moreover, the preparation of the cubic phase state or the cluster state can be experimentally certified by means of quantum tomographic strategies, following for example the scheme recently implemented to verify two-mode entanglement in electro-mechanical systems \cite{kotler2021direct}. In particular, methods for reconstructing the state of a network of harmonic resonators coupled to an auxiliary mode \cite{moore2016quantum} or to a two level system \cite{tufarelli2012reconstructing} have been proposed.


\section{Conclusions and outlook}\label{sec:conclusion}

Continuous-variable systems are convenient for fault-tolerant computation since they naturally offer high-dimensional spaces in which the discrete units of quantum information can be resiliently encoded \cite{chuang1997bosonic, gottesman2001encoding, lund2008fault, menicucci2014fault, michael2016new, fukui2017analog}, as recently proven experimentally in the context of circuit-based quantum computation \cite{ofek2016extending, rosenblum2018cnot, fluhmann2018encoding}. In this respect, the alternative measurement-based approach considered here is promising, thanks to the availability of high threshold schemes \cite{raussendorf2007fault, raussendorf2007topological}. In particular, we have shown that a setting where mechanical oscillators act as the information carriers, rather than photons, provides the advantage that the core ingredients for universal computation ---non-Gaussian cluster states and Gaussian operations--- can be realised unconditionally. This opens the way to deterministic fault-tolerant quantum computation in integrable platforms where linear and quadratic optomechanics-like interactions can be simultaneously achieved. 


\section*{Acknowledgements}

We thank M.~Brunelli, M.~Paternostro, and M. Sillanp\"a\"a for helpful discussions. A.F. and O.H. acknowledge support from the EPSRC project EP/P00282X/1 and the EU Horizon2020 Collaborative Project TEQ (grant agreement nr. 766900). O.H. acknowledges support from the SFI-DfE Investigator programme (grant 15/IA/2864). D.M.~acknowledges the Coordinator Support funds from Queen's University Belfast and project GA20-16577S of the Czech Science Foundation.


\appendix


\section{Validity of the rotating wave approximation}\label{sec:validity-rwa}

The validity of the RWA used in the Hamiltonian derivation of the main text will be justified here. Recall that the Hamiltonian
\begin{multline}\label{eqn:optomech-hamiltonian-lin2}
	H=a^\dagger\sum\limits_{j=1}^N\Big(g_1^{(j)} b_j+g_2^{(j)} b_j^\dagger\\
	+g_3^{(j)} b_j^2+g_4^{(j)} {b_j^\dagger}^2+g_5^{(j)}(b_j^\dagger b_j+b_j b_j^\dagger)\Big)+\text{H.c.}\,,
\end{multline}
is obtained by discarding all time-dependent (counter-rotating) terms and keeping only the resonant ones. The counter-rotating terms may be written as
\begin{widetext}
\begin{equation}
	H_{\text{crt}}=\sum\limits_{j=1}^N\Big[\sum_{\ell=1}^4 H_j^{(\ell)}\EXP{i\ell\Omega_j t}
	+\sum_{k=1,k\ne j}^N H_{j,k}\Big]\ +\text{H.c.}\,,
\end{equation}
\end{widetext}
with the following expressions:
\begin{widetext}
	\begin{align}
		H_j^{(1)}	&=	a^\dagger\left[\alpha_0 G_L^{(j)} b_j^\dagger+\alpha_j^{(-2)} G_L^{(j)} b_j+\alpha_j^{(+1)}G_Q^{(j)} {b_j^\dagger}^2+\sum_{k=1}^N\alpha_j^{(-1)} G_Q^{(k)}\{b_k,b_k^\dagger\}\right]\nonumber\\
					&	+a\left[\alpha_0 G_L^{(j)} b_j+\alpha_j^{(+2)} G_L^{(j)} b_j^\dagger+\alpha_j^{(-1)}G_Q^{(j)} b_j^2+\sum_{k=1}^N\alpha_j^{(+1)} G_Q^{(k)}\{b_k,b_k^\dagger\}\right]^\dagger\,,\\
		H_j^{(2)}	&=	a^\dagger\left[\alpha_0 G_Q^{(j)} {b_j^\dagger}^2+\alpha_j^{(-1)} G_L^{(j)} b_j^\dagger+\sum_{k=1}^N\alpha_j^{(-2)} G_Q^{(k)}\{b_k,b_k^\dagger\}\right]\nonumber\\
					& +a\left[\alpha_0 G_Q^{(j)} b_j^2+\alpha_j^{(+1)} G_L^{(j)} b_j+\sum_{k=1}^N\alpha_j^{(+2)} G_Q^{(k)}\{b_k,b_k^\dagger\}\right]^\dagger\,,\\
		H_j^{(3)}	&=	a^\dagger\left[\alpha_j^{(-2)} G_L^{(j)} b_j^\dagger+\alpha_j^{(-1)} G_Q^{(j)} {b_j^\dagger}^2\right]+a\left[\alpha_j^{(+2)} G_L^{(j)} b_j+\alpha_j^{(+1)} G_Q^{(j)} b_j^2\right]^\dagger\,,\\
		H_j^{(4)}	&=	a^\dagger\left[\alpha_j^{(-2)} G_Q^{(j)} {b_j^\dagger}^2\right]+a\left[\alpha_j^{(+2)} G_Q^{(j)} b_j^2\right]^\dagger\,,\\
		H_{j,k}		&=	\EXP{i(2\Omega_k-\Omega_j)t}\left(a^\dagger\left[\alpha_k^{(-2)}G_L^{(j)} b_j+\alpha_j^{(+1)}G_Q^{(k)} {b_j^\dagger}^2\right]+a\left[\alpha_k^{(+2)}G_L^{(j)} b_j^\dagger+\alpha_j^{(-1)}G_Q^{(k)} b_j^2\right]^\dagger\right)\nonumber\\
		&\quad	+\EXP{i(2\Omega_k+\Omega_j)t}\left(a^\dagger\left[\alpha_k^{(-2)}G_L^{(j)} b_j^\dagger+\alpha_j^{(-1)}G_Q^{(k)} {b_j^\dagger}^2\right]+a\left[\alpha_k^{(+2)}G_L^{(j)} b_j+\alpha_j^{(+1)}G_Q^{(k)} b_j^2\right]^\dagger\right)\nonumber\\
		&\quad	+\EXP{2i(\Omega_k-\Omega_j)t}\left(\alpha_k^{(-2)} a^\dagger+\alpha_k^{(+2),*} a\right)G_Q^{(j)} b_j^2+\EXP{2i(\Omega_k+\Omega_j)t}\left(\alpha_k^{(-2)} a^\dagger+\alpha_k^{(+2),*} a\right)G_Q^{(j)} {b_j^\dagger}^2\nonumber\\
		&\quad	+\EXP{i(\Omega_k+\Omega_j)t}\left(\alpha_k^{(-1)} a^\dagger+\alpha_k^{(+1),*} a\right)G_L^{(j)} b_j^\dagger+\EXP{i(\Omega_k-\Omega_j)t}\left(\alpha_k^{(-1)} a^\dagger+\alpha_k^{(+1),*} a\right)G_L^{(j)} b_j\,.
	\end{align}
\end{widetext}

Now we can state the necessary conditions to safely neglect the counter-rotating terms. For the RWA to be valid, the following constraints must be met:
\begin{align}
	\left|\alpha_0G_{L,Q}^{(j)}\right|			&\ll	\Omega_j\,,\\
	\left|\alpha_j^{(\pm1)}G_{L,Q}^{(k)}\right|	&\ll	\Omega_j\,,\\
	\left|\alpha_j^{(\pm2)}G_{L,Q}^{(k)}\right|	&\ll	\Omega_j\,.
\end{align}
	
We study the validity of the RWA in more details for the interesting case of the preparation of the cubic phase state of a mechanical oscillator. The system consists of a cavity and one mechanical oscillator ($N=1$). The full Hamiltonian of the system is again
\begin{equation}
	H=H_{\text{RWA}}+H_{\text{crt}}\,,
\end{equation}
with :
\begin{widetext}
	\begin{align}
		H_{\text{RWA}}	&=	a^\dagger\left(g_1b+b_2 b^\dagger+g_3 b^2+g_4{b^\dagger}^2+g_5\{b,b^\dagger\}\right)\ +\text{H.c.}\,,\\
		H_{\text{crt}}	&=	\sum_{\ell=1}^4 H^{(\ell)}\EXP{i\ell\Omega t}\ +\text{H.c.}\,,
	\end{align}
\end{widetext}
and $H^{(\ell)}$ given by
\begin{widetext}
	\begin{align}
		H^{(1)}	&=	R(g_3 a^\dagger+g_4^* a)b+R(g_5 a^\dagger+g_5^* a)b^\dagger+R^{-1}(g_2 a^\dagger+g_1^* a){b^\dagger}^2+R^{-1}(g_1 a^\dagger+g_2^* a)\{b,b^\dagger\}\,,\\
		H^{(2)}	&=	(g_1 a^\dagger+g_2^* a)b^\dagger+(g_5 a^\dagger+g_5^* a){b^\dagger}^2+(g_3 a^\dagger+g_4^* a)\{b,b^\dagger\}\,,\\
		H^{(3)}	&=	R(g_3 a^\dagger+g_4^* a)b^\dagger+R^{-1}(g_1 a^\dagger+g_2^* a){b^\dagger}^2\,,\\
		H^{(4)}	&=	(g_3 a^\dagger+g_4^* a){b^\dagger}^2\,,
	\end{align}
\end{widetext}
where we defined $R\equiv\frac{G_L}{G_Q}$ the ratio between the bare linear and quadratic optomechanical couplings. Therefore, the necessary conditions for the validity of the RWA are
\begin{multline}
	|g_j|,|R g_\mu|,|R^{-1} g_\nu|\ll\Omega\,,\\
	(j=1,\ldots,5,\ \mu=3,4,5,\ \nu=1,2)\,.
\end{multline}

In particular, for the cubic phase state; $g_2=-r g_1$, $g_3=g_4=g_5=\frac{-3i}{2\sqrt2}\ (1+r)\gamma g_1$ ($0\le r<1$ and $\gamma$ real), these latter conditions translate to
\begin{equation}
	|g_1|,\ |R g_1|,\ |R^{-1} g_1|\ll\Omega\,.
\end{equation}
	
\begin{figure}[hbt]
		\includegraphics[scale=0.55]{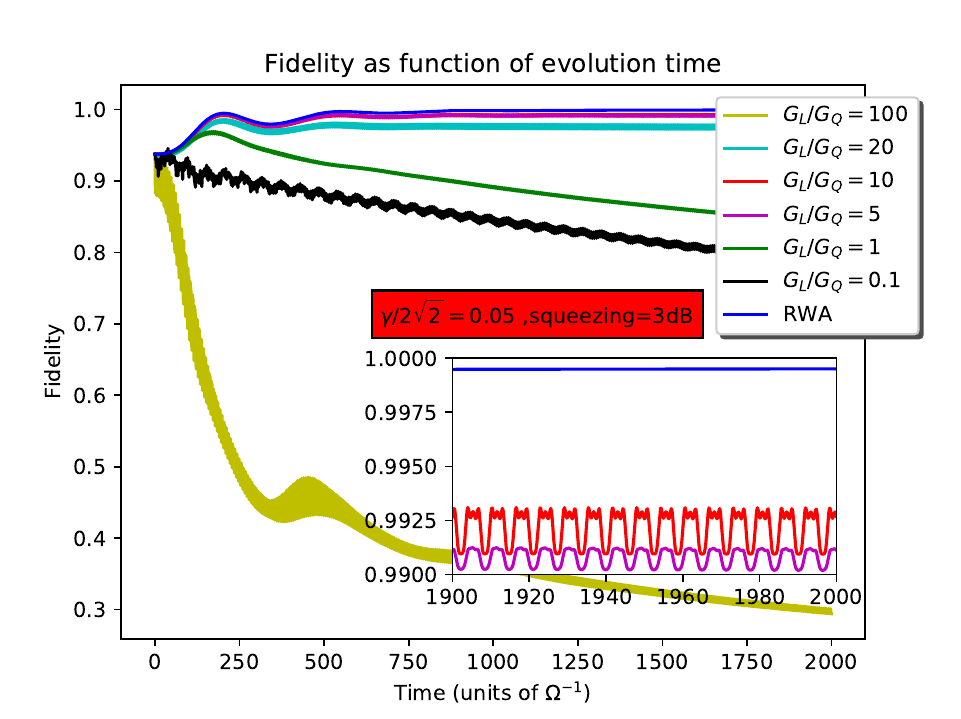}
		\caption{\label{fig:fidelity-counter-rotating}Fidelity of the system's state with the cubic phase state as function of time. The system state is obtained by solving the dynamics of the cavity-mechanical oscillator with and without the RWA. We used the parameters $\gamma/2\sqrt2=0.05$, $r=0.33$ ($3\dB$ squeezing) and $g_1=\kappa=10^{-2}\ \Omega$.}
	\end{figure}

In the following we quantify the effect of the counter rotating terms on the steady state of the dissipative dynamics. For this, we use the Uhlman fidelity defined as \cite{Uhlmann:76,jozsa1994fidelity}:
\begin{equation}
	F(\gamma,s)=\sqrt{\braket{\gamma,s|\rho_\text{full}(t)|\gamma,s}}\,,
\end{equation}
	where $\rho_\text{full}(t)$ is the density operator of the system at time $t$ when considering the full Hamiltonian, $H_{\mathrm{RWA}}+H_{\mathrm{crt}}$. We calculate $\rho_\text{full}(t)$ by solving the master equation for big enough Hilbert space and plot the fidelity $F\left(0.05\times 2\sqrt2,s(0.33)\right)$ as function of time (see \reffig{fig:fidelity-counter-rotating}). We see that one can reach fidelity $>0.99$ in some regimes. Namely,  for the used values and when the ratio $R=\frac{G_L}{G_Q}$ is between 5 and 10, the validity of the RWA is justified.
	
\section{Stability analysis}\label{sec:stability}
		
Here we will give a detailed analysis of the stability of the optomechanical system described by the Hamiltonian \Refeq{eqn:optomech-hamiltonian-lin2} for one mechanical oscillator ($N=1$). The Langevin-Heisenberg equations for the quantum fluctuations are:
\begin{equation}\label{eqn:eqs-motion-fluct}
	\dot{\mathcal{U}}=A\ \mathcal{U}+B+\mathcal{N}\,,
\end{equation}
where $\mathcal{U}$ is the operator-valued vector defined as $\mathcal{U}=(x,y,q,p)^\top$, with $x=(a+a^\dagger)/\sqrt2$ and $y=(a-a^\dagger)/\sqrt2 i$ the cavity field quadratures, $\mathcal{N}=(x_{\text{in}},y_{\text{in}},0,\xi)^\top$ is the vector of noise operators, and the matrix $A$ and the vector $B$ are given by:
\begin{widetext}
	\begin{equation}
		A=
		\left(
		\begin{array}{cccc}
			\frac{-\kappa}{2}	&	0					&	I_1		&	R_2\\
			0					&	\frac{-\kappa}{2}	&	-R_1	&	I_2\\
			-I_2				&	R_2					&	0		&	0\\
			-R_1				&	-I_1				&	0		&	-\Gamma
		\end{array}
		\right)
		\qquad,\qquad
		B=
		\left(
		\begin{array}{c}
			I_3 q^2+I_4 p^2+R_5(qp+pq)\\
			-R_3 q^2-R_4 p^2+I_5(qp+pq)\\
			2x(R_4 p-I_5 q)+2y(I_4 p+R_5 q)\\
			2x(-R_3 q+I_5 p)-2y(I_3 q+R_5 p)
		\end{array}
		\right)\,,
	\end{equation}
\end{widetext}
where $R_k$ and $I_k$, $k=1,\ldots,5$, are defined as follows:
\begin{widetext}
	\begin{align}
		R_1&=\Re(g_1+g_2)						& I_1&=\Im(g_1+g_2)\,,\\
		R_2&=\Re(g_1-g_2)						& I_2&=\Im(g_1-g_2)\,,\\
		R_3&=\frac{1}{\sqrt2}\Re(g_3+g_4+2g_5)	& I_3&=\frac{1}{\sqrt2}\Im(g_3+g_4+2g_5)\,,\\
		R_4&=\frac{-1}{\sqrt2}\Re(g_3+g_4-2g_5)	& I_4&=\frac{-1}{\sqrt2}\Im(g_3+g_4-2g_5)\,,\\
		R_5&=\frac{1}{\sqrt2}\Re(g_3-g_4)		& I_5&=\frac{1}{\sqrt2}\Im(g_3-g_4)\,.
	\end{align}
\end{widetext}
	The system given by \refeq{eqn:eqs-motion-fluct} is stable if the linear part is stable \cite{woolley2008nanomechanical,jiang2016controllable,mikkelsen2017optomechanics}. This is equivalent to $A$ being a Hurwitz matrix, \ie all eigenvalues have negative real part. In fact, applying the Routh-Hurwitz criterion \cite{dejesus1987routh} we find the following stability condition:
	\begin{equation}
		R_1R_2+I_1I_2>0\qquad\Leftrightarrow\qquad|g_1|>|g_2|\,,
	\end{equation}
	\ie the driving amplitude at blue side band is smaller than that at the red side band. For the cubic phase state, we have $g_2=-r g_1$. Therefore the system is always stable as long as $|r|<1$, which is always the case, since $r=\frac{s^2-1}{s^2+1}$ and $s\ge 1$.

\section{Time scale to reach the target cluster state}\label{sec:time-scale-cluster}
	At each step of the switching protocol, the Hamiltonian is set to
	\begin{equation}\label{eqn:hamiltonian-a-d-ell}
		H_\ell=\beta_\ell(a^\dagger d_\ell+a d_\ell^\dagger)\,,
	\end{equation}
	where $\beta_\ell$ is proportional to the driving power. Since the driving power may differ in every step we attach the subscript $\ell$ to $\beta$.
	
	The dynamics of the system, at step $\ell$, is governed by the master equation
    \begin{equation}
        \dot{\rho}_\ell=-i[H_\ell,\rho_\ell]+\kappa\left(a\rho_\ell a^\dagger-\frac12 a^\dagger a\rho_\ell-\frac12 \rho_\ell a^\dagger a\right)\,.
    \end{equation}
    The system will reach the steady state (of step $\ell$) in a time scale given by \cite{houhou2015generation}
    \begin{equation}
        \tau_\ell=\frac{4}{\kappa\Re\left(1-\sqrt{1-\left(\frac{4\beta_\ell}{\kappa}\right)^2}\right)}\,.
    \end{equation}
    If the target state has size $N$ (\ie $N$ mechanical oscillators), then the switching scheme involves $N$ steps, and the time scale to prepare the cluster state is
    \begin{equation}
        \tau(N)=\sum_{\ell=1}^N\tau_\ell\,.
    \end{equation}
    In the simplest setting where all driving powers are identical, we set $\beta_\ell\equiv\beta$ constant in all steps and obtain:
    \begin{equation}
        \tau(N)=\frac{4N}{\kappa\Re\left(1-\sqrt{1-\left(\frac{4\beta}{\kappa}\right)^2}\right)}\,.
    \end{equation}
    This shows that the timescale to prepare the cluster state grows linearly with the number of nodes. For example, if we had chosen $\kappa=10\beta$, then the time scale in units of $\beta$ is found to be $\tau(2)=9.58$. The scheme is most effective when this timescale is less than the rethermalisation time of the system, \ie $\frac{1}{\tau(N)}<\bar{n}\Gamma$, where $\bar{n}\Gamma$ is the worst case for the collection of oscillators.

\section{Two-mode non-Gaussian cluster state}\label{sec:two-modes-cluster}

\begin{figure}[hbt]
	\begin{center}
		\includegraphics[scale=1.0]{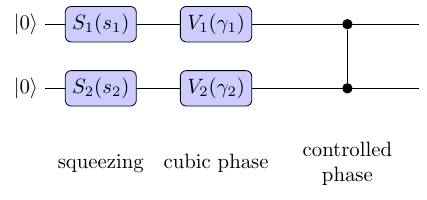}
	\end{center}
	\caption{\label{fig:two-modes-cluster}Quantum circuit for the two modes non-Gaussian cluster state. Quantum gates $S_1$ and $S_2$ ($V_1$ and $V_2$) are the one-mode squeezing (cubic phase) for modes~1 and~2 respectively.}
\end{figure}

We demonstrate the generation of two-mode non-Gaussian cluster state using the protocol introduced so far. We choose the target cluster (see \reffig{fig:two-modes-cluster}) to be a squeezed state (with parameters $s_1$ and $\gamma_1=0$) coupled to a cubic phase state (with parameters $s_2$ and $\gamma_2$).

First we focus on the noiseless case, \ie no mechanical dumping. Since we have two mechanical modes than preparing the target cluster involves two steps. Starting from the vacuum state of the two mechanical oscillators, we set the driving amplitudes such that the system's Hamiltonian is
\begin{widetext}
	\begin{equation}
		H_1=\frac{\beta}{2} a^\dagger\left[\left(s_1+\frac{1}{s_1}\right)b_1-\left(s_1-\frac{1}{s_1}\right)b_1^\dagger-i s_1\left(b_2+b_2^\dagger\right)-\frac{3i\gamma_1 s_1}{\sqrt2}\left(b_1+b_1^\dagger\right)^2\right]+\text{H.c.}\,,
	\end{equation}
\end{widetext}
and we wait for sufficient time to reach the steady state. Then we set the amplitudes so that the Hamiltonian is
\begin{widetext}
	\begin{equation}
		H_2=\frac{\beta}{2} a^\dagger\left[-i s_2\left(b_1+b_1^\dagger\right)+\left(s_2+\frac{1}{s_2}\right)b_2-\left(s_2-\frac{1}{s_2}\right)b_2^\dagger-\frac{3i\gamma_2 s_2}{\sqrt2}\left(b_2+b_2^\dagger\right)^2\right]+\text{H.c.}\,,
	\end{equation}
\end{widetext}
and we wait again for sufficient time to reach the steady state of the system. In main text we showed a plot (\reffig{fig:TwoModeFidelity}) showing numerical confirmation that the system reaches the target cluster as a steady state of the system.
		
Now we turn our focus to the noisy case where the dynamics suffers from the (unwanted) coupling of the mechanical oscillators with their thermal baths at finite temperature. We assess the quality of the generated cluster state by the two steps protocol detailed above. We consider that the two mechanical oscillators are initially in thermal equilibrium with their baths with mean phonon occupations $\bar n_1$ and $\bar n_2$ for the first and second modes respectively. Without loss of generality we assume same mechanical damping rate $\Gamma$ for both oscillators. The fidelity of the system was plotted in \reffig{fig:TwoModeFidelity} in main text for $\bar n_1=10$, $\bar n_2=1$ and $\Gamma=10^{-4}\beta$. We notice that there is a gap between the curves corresponding to the noisy and noiseless cases, and this is mainly due to the fact that our protocol is valid when the initial state of the mechanical oscillators is the vacuum. And since the mechanical oscillators here are initially in thermal state, then we will expect that the fidelity will follow a different path from that corresponding to the system being initially in vacuum. This is true even if the noise is disregarded during the switching protocol. To be able to assess the robustness of our protocol against the effects of the mechanical noise, we suggest cooling down the mechanical oscillators before starting the switching protocol. The cooling process is realised by exploiting the red side band cooling of each mechanical oscillator individually \cite{lei2016quantum}: We drive the system with one field addressing one mechanical oscillator only. The implemented Hamiltonian writes:
\begin{equation}
	H_j^{\text{cool}}=\beta a^\dagger b_j+\text{H.c.}\,,
\end{equation}
for $j=1,\ 2$. Therefore, our protocol involves four steps: two steps for cooling the first then second mechanical oscillators, and two steps for the preparation of the target cluster state as explained above. Hence, when pre-cooling the oscillators, the target cluster state is obtained with higher fidelity than before.

\section{Three-mode non-Gaussian linear cluster state}\label{sec:three-modes-cluster}
\begin{figure}[hbt]
		\begin{center}
			\includegraphics[scale=0.6]{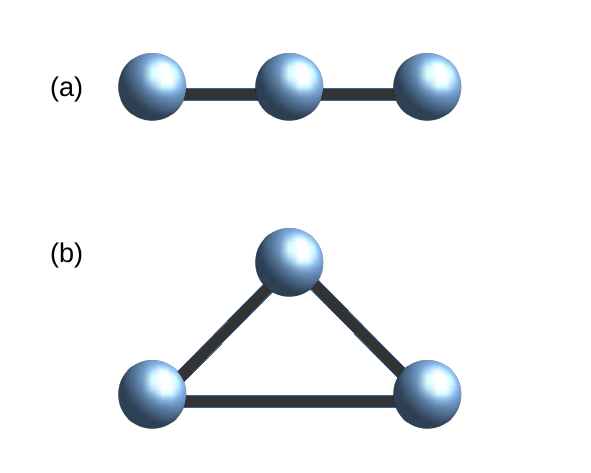}
		\end{center}
		\caption{\label{fig:three-mode-cluster-geometries}Possible geometries of a three-mode cluster state. (a)~Linear cluster. (b)~Circular cluster.}
	\end{figure}

\begin{figure}[hbt]
		\begin{center}
			\includegraphics[scale=0.7]{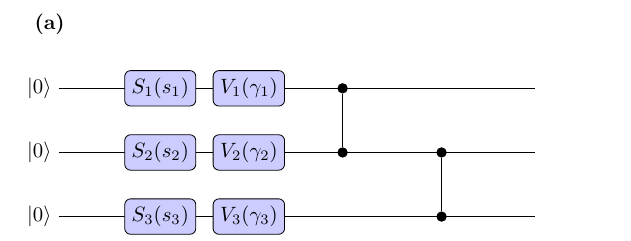}\\[0.5cm]
			\includegraphics[scale=0.7]{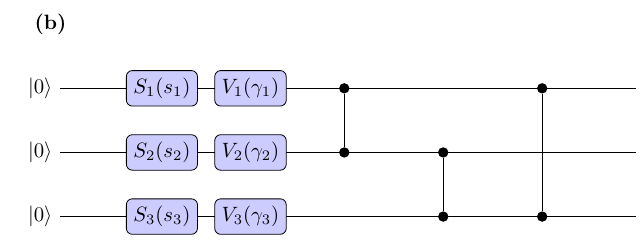}
		\end{center}
		\caption{\label{fig:three-mode-cluster-circtuits}Quantum circuit for generating a linear (a) and circular (b) three-mode cluster state.}
	\end{figure}

	In this appendix, we show the needed steps to prepare a non-Gaussian three-mode cluster state. In contrast to the case of two-mode cluster, there are different geometries for the three-mode cluster. Namely, the linear and circular geometries, see \reffig{fig:three-mode-cluster-geometries}. The canonical way of preparing the three-mode cluster state is depicted in the quantum circuits of \reffig{fig:three-mode-cluster-circtuits}.

	Using our scheme given in \refsec{sec:non-gaussian-cluster}, we need three steps. In each step the Hamiltonian of the system is set as the following (we used equations~\Refeq{eqn:H-cluster} and~\Refeq{eqn:d-ell}). For the linear cluster we have :
	\begin{widetext}
		\begin{align}
			H_1^\text{linear}	&=	\frac{\beta}{2} a^\dagger\left[\left(s_1+\frac{1}{s_1}\right)b_1-\left(s_1-\frac{1}{s_1}\right)b_1^\dagger-i s_1\left(b_2+b_2^\dagger\right)-\frac{3i\gamma_1 s_1}{\sqrt2}\left(b_1+b_1^\dagger\right)^2\right]+\text{H.c.}\,,\\
			H_2^\text{linear}	&=	\frac{\beta}{2} a^\dagger\left[-i s_2\left(b_1+b_1^\dagger+b_3+b_3^\dagger\right)+\left(s_2+\frac{1}{s_2}\right)b_2-\left(s_2-\frac{1}{s_2}\right)b_2^\dagger-\frac{3i\gamma_2 s_2}{\sqrt2}\left(b_2+b_2^\dagger\right)^2\right]+\text{H.c.}\,,\\
			H_3^\text{linear}	&=	\frac{\beta}{2} a^\dagger\left[-i s_3\left(b_2+b_2^\dagger\right)+\left(s_3+\frac{1}{s_3}\right)b_3-\left(s_3-\frac{1}{s_3}\right)b_3^\dagger-\frac{3i\gamma_3 s_3}{\sqrt2}\left(b_3+b_3^\dagger\right)^2\right]+\text{H.c.}\,,
		\end{align}
	\end{widetext}
	and for the circular cluster the Hamiltonian formulae are similar to the above with minor modifications due to the extra coupling between first and third modes:
	\begin{align}
		H_1^\text{circular}	&=	H_1^\text{linear}\ +\ \frac{i s_1\beta}{2} \left(a-a^\dagger\right)\left(b_3+b_3^\dagger\right)\,,\\
		H_2^\text{circular}	&=	H_2^\text{linear}\,,\\
		H_3^\text{circular}	&=	H_3^\text{linear}\ +\ \frac{i s_3\beta}{2} \left(a-a^\dagger\right)\left(b_1+b_1^\dagger\right)\,.
	\end{align}

\begin{figure}[hb]
		\includegraphics[scale=0.6]{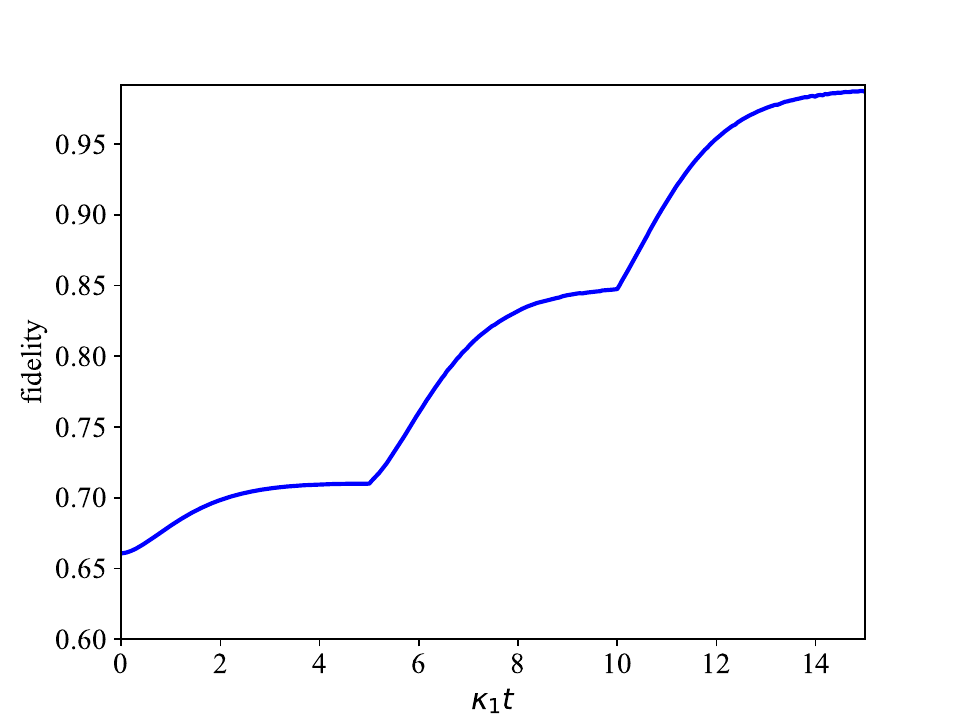}
		\caption{Variation of the fidelity between the state of the system at time $t$ and the target cluster state (three-mode cluster with linear geometry) with time. We used the following parameters: $s_1=s_2=s_3\approx 1.41$ (3~dB squeezing), $\gamma_1=0.1$, $\gamma_2=\gamma_3=0$.}
		\label{fig:three-mode-cluster-fidelity}
	\end{figure}

	In \reffig{fig:three-mode-cluster-fidelity} we plot the fidelity between a linear three-mode cluster (see the figure for the used parameters) and the time evolution of the system's state during the two switching steps. As expected, the system reaches a steady state in every switching step, witnessed by a constant fidelity with time. More interestingly, the system's steady state at the final step is exactly the target cluster state, and this is clear from the fact that the fidelity reached stationary value of one.

	Needless to say, the simulation of other cluster states with more modes becomes computationally very difficult due to the exponential growth of computation resources needed to perform the simulations.

\section{Mechanical noise effects on the preparation of the cluster states}\label{sec:noise-cluster}

In this section, we discuss the notorious effects of the mechanical thermal noise on the quality of the cluster states obtained from the switching protocol. Our system involves $N$ mechanical oscillators interacting with a common cavity mode. Recall that the full Hamiltonian is given by
\begin{multline}\label{eqn:hamiltonian-a-b}
	\mathcal{H}=a^\dagger\sum_j\Big(g^{(j)}_1 b_j+g^{(j)}_2 b_j^\dagger\\
	+g^{(j)}_3 b_j^2+g^{(j)}_4 {b_j^\dagger}^2+g^{(j)}_5\{b_j,b_j^\dagger\}\Big)+\text{H.c.}\,,
\end{multline}
and the dynamics obeys the master equation:
\begin{widetext}
	\begin{equation}\label{eqn:master-eqn-noise}
		\frac{\diff\rho(t)}{\diff t}=-i[\mathcal{H},\rho(t)]+\kappa D[a]\rho(t)+\sum\limits_{j=1}^N\ \Big(\Gamma_j(\bar n_j+1)D[b_j]\rho(t)+\Gamma_j \bar n_j D[b_j^\dagger]\rho(t)\Big) \,,
	\end{equation}
\end{widetext}
where $\Gamma_j$ and $\bar n_j$ are, respectively, the damping rate and the mean-phonon number corresponding to the mechanical oscillator $j$.

In the following simulations, we consider a system of two mechanical oscillators and assume they are initially in thermal equilibrium with their respective baths. After cooling down the two oscillators, we apply the switching protocol (see the main text and Appendix~\ref{sec:two-modes-cluster}) and calculate the final fidelity at the steady state. Due to the computational difficulty of simulating this system, we consider a regime where the cavity mode can be adiabatically eliminated. Namely, we consider that the linear ($g_{1,2}^j$) and quadratic ($g_{3,4,5}^j$) optomechanical couplings are much less than the cavity decay rate ($\kappa$). The system dynamics is now described by the following master equation \cite{Gardiner:00,wiseman1993quantum,woolley2014two,brunelli2018unconditional}
\begin{widetext}
	\begin{multline}\label{eqn:master-eqn-noise-effective}
		\frac{\diff\rho(t)}{\diff t}=\kappa_1 D\left[\sum_{j=1}^Ng^{(j)}_1 b_j+g^{(j)}_2 b_j^\dagger+g^{(j)}_3 b_j^2+g^{(j)}_4 {b_j^\dagger}^2+g^{(j)}_5\{b_j,b_j^\dagger\}\right]\rho(t)\\
		+\sum\limits_{j=1}^N\ \Gamma_j(\bar n_j+1)D[b_j]\rho(t)+\Gamma_j \bar n_j D[b_j^\dagger]\rho(t) \,,
	\end{multline}
\end{widetext}
with the effective decay rate $\kappa_1=\frac{4\beta^2}{\kappa}$. For simplicity we set $\Gamma_1=\cdots=\Gamma_N$.

\begin{figure}[H]
	\begin{center}
		\includegraphics[scale=0.6]{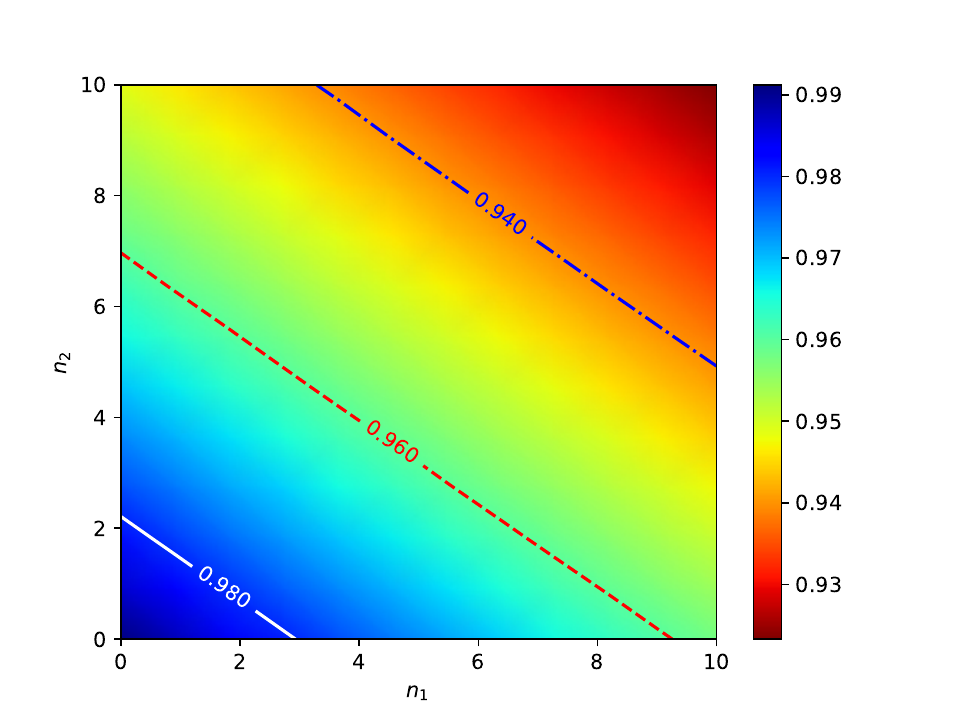}
	\end{center}
	\caption{Final fidelity as function of temperatures for the two mechanical oscillators. The used parameters are: $s_1=s_2\approx 1.41$ (3~dB squeezing), $\gamma_1=0.1$, $\gamma_2=0$, $\Gamma_1=\Gamma_2=10^{-3}\kappa_1$.}
	\label{fig:final-fidelity-temperature}
\end{figure}

By varying the temperature for the two oscillators, we obtained a contour plot, see \reffig{fig:final-fidelity-temperature}. As one would expect, the presence of mechanical noise has a deleterious effect on the prepared cluster state; the greater the temperature of the two mechanical oscillators the larger the deviation of the steady state from the ideal target cluster. This deviation is the result of two things; firstly, the initial state is no longer the vacuum (our protocol requires the vacuum as the initial state). To counter the effect of a non-ideal initial state, we perform the cooling stage for all the mechanical oscillators. Secondly, the presence of mechanical coupling to the thermal baths will further affect the quality of the cluster at the steady state. In fact, it is better not to wait for a very long time to reach the steady state, but one may consider shorter times per switching step that are less than the decoherence time due to thermal effects \cite{houhou2015generation}.

\section{Cubic phase gate}\label{sec:cubic-phase-gate}
	
We consider the two-node non-Gaussian cluster of \reffig{fig:two-modes-cluster}. We perform a momentum measurement on the input squeezed state, which results (up to a distortion due to finite squeezing) in the output given by:
\begin{equation}\label{eqn:two-mode-output}
	\ket{\phi'}=X(m)P(3\gamma m)Z(3\gamma m^2)F\EXP{-i\gamma p^3}\ket{\phi}\,.
\end{equation}

\begin{figure}[H]
	\begin{center}
		\includegraphics[scale=0.5]{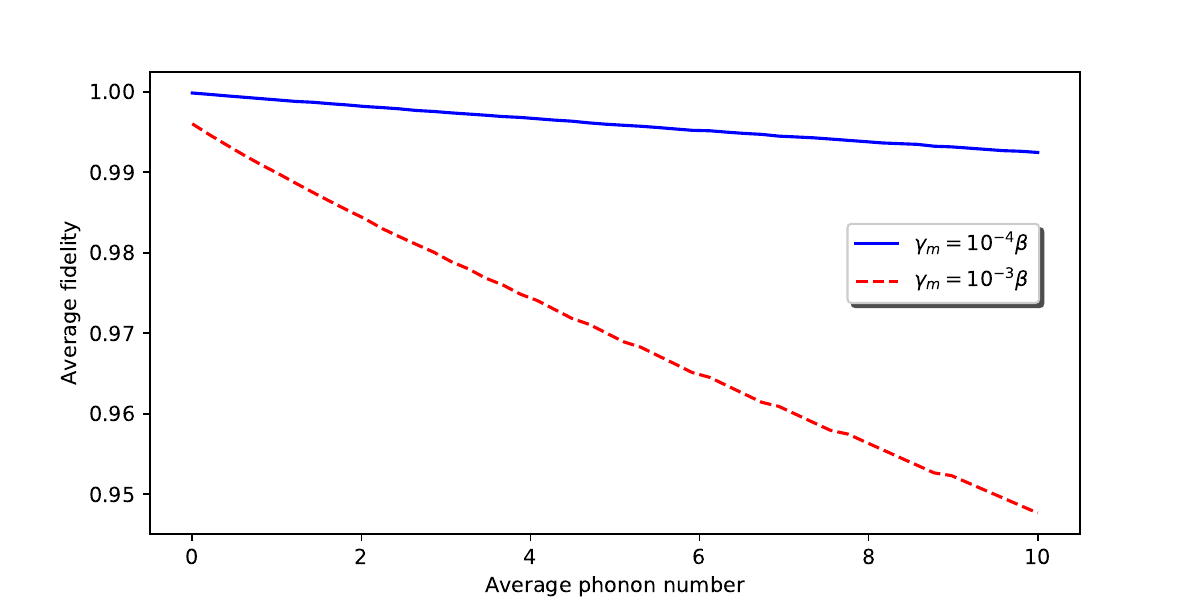}
	\end{center}
	\caption{Fidelity of the output state with the cubic phase gate target state averaged over the many measurements. On average the operation produces a state with high fidelity to the target, with large temperatures and high damping rates leading to smaller fidelities. See the text for the used parameters.}
	\label{fig:avgfidelity}
\end{figure}

We assume the same conditions as in the generation of the non-Gaussian cluster and further assume the capability to make a projective measurement on the input node (cf \cite{darren2017arbitrary}). The fidelity of the output state with \refeq{eqn:two-mode-output} is analysed in \reffig{fig:avgfidelity}. Since the output depends on the measurement result, which is random, we examine the fidelity on average over many measurement results. The scheme proves effective on average with decreasing success as the temperature increases.

We should mention that in all the simulations carried out in our paper, including the measurement simulation, we approximated the {\em ideal} continuous system with a {\em discrete} one by truncating the dimension of the Hilbert space in the Fock basis. Consequently, the spectrum of the momentum observable becomes finite and discrete. Therefore, measuring the momentum $\hat{P}$ will always give a result that belongs to this discrete spectrum. In particular, the measurement simulations shown in \reffig{fig:avgfidelity} have a finite resolution which intrinsically constitutes a binning process. In other words, the discreteness of the spectrum of $\hat{P}$ is essentially a form of binning forced on us by the structure of the simulation. Therefore the measurement leading to the results in \reffig{fig:avgfidelity} has an intrinsic non-zero measurement width. Given that this particularly coarse resolution is successful, we expect that a more realistic scenario with narrower widths will be even more successful.


\newcommand{\noop}[1]{}

\end{document}